\documentclass[runningheads,anonymous=true,review]{llncs}

\usepackage{graphicx}
\usepackage{array}
\usepackage{amssymb}
\usepackage{amsmath}
\usepackage{multirow}
\usepackage{arydshln}
\usepackage{diagbox}
\usepackage{enumitem}
\usepackage{adjustbox}
\usepackage[para]{footmisc}
\usepackage{stfloats}
\usepackage{cleveref}
\usepackage{makecell}
\usepackage[dvipsnames]{xcolor}
\usepackage[normalem]{ulem}
\usepackage{mathtools}
\usepackage{booktabs}
\usepackage{tablefootnote}

\usepackage{balance}
\usepackage{threeparttable}
\usepackage[numbers,sort&compress]{natbib}

\usepackage[group-minimum-digits=3,group-separator={{,}}]{siunitx}

\newcommand{\maxscore}{\method{Max\-Score}}

\newcommand{\method}[1]{{\small\sf{#1}}}

\newcommand{\myparagraph}[1]{\paragraph{\bf{#1}.}}

\usepackage{xcolor}

\newcommand{\collection}[1]{\method{#1}}
\newcommand{\govtwo}{\collection{Gov2}}

\newcommand{\msmarcovone}{\collection{MS MARCO}}
\newcommand{\msm}{\collection{\footnotesize MSM}}

\newcommand{\BM}{\method{BM25}}

\newcommand{\DeepImpact}{{\method{DeepImpact}}}
\newcommand{\uniCOIL}{{\method{uniCOIL}}}
\newcommand{\uCOIL}{{\method{uCOIL}}}

\newcommand{\DocTQuery}{{\method{DocT5Query}}}
\newcommand{\SPLADE}{{\method{SPLADE}}}
\newcommand{\SPL}{{\method{SPL}}}
\newcommand{\TILDE}{{\method{TILDE}}}
\newcommand{\TILDEA}{{\method{TILDE-A}}}
\newcommand{\TILDEACSV}{{\method{TILDE-AUG-CSV}}}

\begin{document}
\title{Improved Learned Sparse Retrieval with Corpus-Specific Vocabularies}
%
%
\author{Puxuan Yu\inst{1} \and
Antonio Mallia\inst{2}\thanks{Work partly done while working at Amazon Alexa.} \and
Matthias Petri\inst{3}}
%
%
\institute{University of Massachusetts Amherst, USA \and
Pinecone, Italy \and
Amazon AGI, USA\\
\email{pxyu@cs.umass.edu}, \email{antonio@pinecone.io}, \email{mkp@amazon.com}
}
%
\maketitle              
\begin{abstract}
We explore leveraging corpus-specific vocabularies that improve both efficiency and effectiveness of learned sparse retrieval systems. We find that pre-training the underlying BERT model on the target corpus, specifically targeting different vocabulary sizes incorporated into the document expansion process, improves retrieval quality by up to 12\% while in some scenarios decreasing latency by up to 50\%. Our experiments show that adopting corpus-specific vocabulary and increasing vocabulary size decreases average postings list length which in turn reduces latency. Ablation studies show interesting interactions between custom vocabularies, document expansion techniques, and sparsification objectives of sparse models. Both effectiveness and efficiency improvements transfer to different retrieval approaches such as uniCOIL and SPLADE and offer a simple yet effective approach to providing new efficiency-effectiveness trade-offs for learned sparse retrieval systems.

\keywords{Learned sparse retrieval \and Language model vocabulary.}
\end{abstract}
\section{Introduction}
Sparse term representations such as 
\SPLADE~\cite{Formal-etal-2021-splade},
\TILDE~\cite{tildev1} or \uniCOIL~\cite{coil,arxiv21lm} establish 
competitive retrieval performance using existing sparse 
retrieval techniques underpinned by standard
{\emph{inverted indexes}} data structures~\cite{zm06-csurv}. The inverted index has been 
optimized to be highly scalable, cost-efficient, update-able in real-time, and continue
to be one of the core first-stage retrieval components in most commercial search
systems today.

\begin{figure}[t]
\includegraphics[width=0.9\columnwidth]{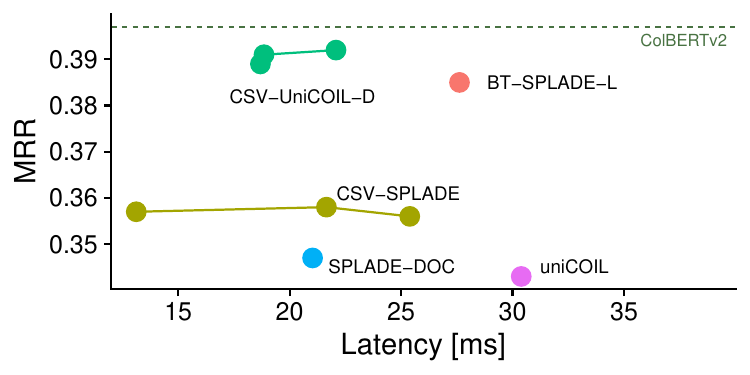}
\caption{Latency and effectiveness improvements achieved by leveraging Corpus Specific Vocabularies (\method{CSV}) (with different vocabulary sizes) compared to baseline learned sparse retrieval models.}
\label{fig:teaser}
\end{figure}

One of the key distinctions of state-of-the-art learned sparse representations compared to traditional ranking functions such as \BM~\cite{rz09fntir} is the tight integration between the vocabulary of the inverted index and the one of the model producing term importance representations for each
document. While \BM~based inverted indexes contain potentially millions of unique tokens,
learned sparse indexes generally restrict the vocabulary to tokens occurring in the underlying BERT~\cite{Devlin-etal-2019-bert} vocabulary. This vocabulary is usually restricted to say 30,000 entries to improve model efficiency.

While some work has elucidated the link between term score distribution and learned sparse representations~\cite{mmmp22-femnlp,mmst2022gt}, in this work we explore the relationship between vocabulary selection, retrieval quality, and runtime efficiency of learned sparse representations. 

\myparagraph{\hspace{-0.45cm} Contribution}

This work provides the following contributions:

\begin{itemize}[leftmargin=*]
    \item We show the benefit of creating corpus-specific vocabularies to pre-train underlying language models to retrieval quality.
    \item We explore trade-offs between vocabulary size, pre-training time, document expansion, and effectiveness improvements.
    \item We demonstrate that our approach is applicable to many state-of-the-art techniques such as \SPLADE, and \uniCOIL. 
    \item We propose a corpus-specific modification to \TILDE{} document expansion that leverages custom vocabularies as well as augmentation of hard negatives at training time. 
    \item We analyze improvements in retrieval latency resulting from large corpus-specific vocabularies.
\end{itemize}

\noindent Overall our proposed approach is simple and offers new performance trade-offs for different learned sparse models (see Figure~\ref{fig:teaser}).


\section{Background and Related Work}

\myparagraph{Learned Sparse Models} Usage of pre-trained contextualized language
models (LMs) has resulted in improvements to search effectiveness, albeit
with higher retrieval costs than traditional lexical models~\cite{mtl21-arxiv}. 
While models such as \BM~leverage term frequency statistics to estimate term importance
in a document, LMs can be leveraged to learn the importance of a term in a document by 
directly optimizing for the actual retrieval task. These term importance scores form the basis of many {\emph{learned sparse}} retrieval techniques that still
leverage the inverted index for query processing. Such models 
include \SPLADE~\cite{Formal-etal-2021-splade}, \TILDE~\cite{tildev1}, \DeepImpact~\cite{sigir21mkts} or \uniCOIL~\cite{coil,arxiv21lm} which differ in their handling of document and query
processing, vocabulary selection, and training objective but offer state-of-the-art 
retrieval performance while providing different efficiency and effectiveness trade-offs.

\myparagraph{Pre-training} Pre-training refers to allowing a model to learn general language representations by performing tasks such as Masked Language Modeling (MLM) on large text corpora. In the search setting,
techniques such as coCondenser~\cite{cocondenser} provide additional search-specific pre-training tasks to improve the performance of LMs on the actual retrieval task. Such pre-training objectives
may operate on the target retrieval corpus, or larger potentially out-of-domain text corpora.
Recent work has explored the relationship of vocabulary size in standard-pretraining arrangements~\cite{feng2022pretraining} as well as the notion of rare-terms in pre-training requiring
special consideration~\cite{yu2022dictbert}.

\myparagraph{Document Expansion} To mitigate the vocabulary mismatch problem~{\cite{Zhao-2012-vocabmismatch}}, learned sparse representations perform document expansion to augment the document with potentially relevant (future query) tokens. The \DocTQuery~\cite{doct5query} technique augments documents with tokens by appending generated queries from the source document, while \TILDE~\cite{tildev1} directly optimizes for both term importance estimation and document expansion.

\myparagraph{Inverted Index and Dynamic Pruning} 

The {\emph{inverted index}} stores one {\emph{postings list}} for each unique term $t$ produced by a ranking model. Each postings list comprises a sequence of the document ID and corresponding term importance score pairs~{\cite{zm06-csurv,pv21-compsurv}}. During query
processing, the posting lists of all query terms are processed
to retrieve the top-$k$ highest-scoring documents. Query processing algorithms such as the {\maxscore}~\cite{tf95-ipm} or \method{BlockMaxWand}~\cite{ds11-sigir} dynamic
pruning mechanisms enable skipping of large sections of postings lists. However, a relationship still exists between the length of each postings list and overall query latency~\cite{wsdm-disjunctive22}.

\section{Corpus-specific Vocabularies}
\label{sec:csv}

\begin{figure}[t]
\includegraphics[width=\textwidth]{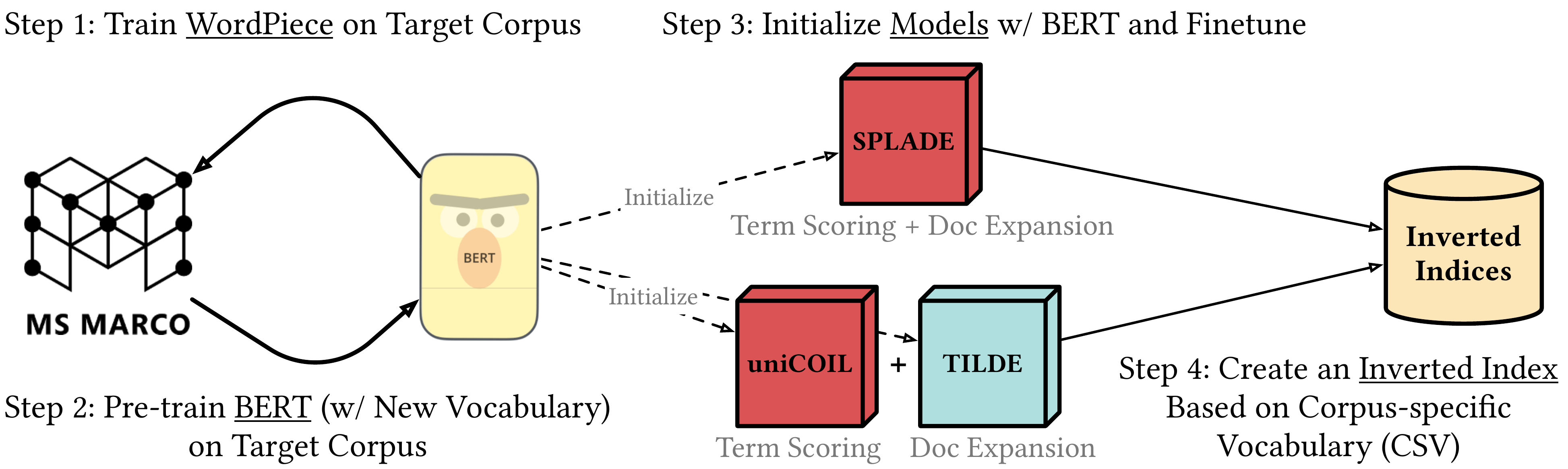}
\caption{A high-level overview of the workflow described in this work. As the vocabulary of the language model is learned on the target retrieval corpus, and that the sparse retrieval models (e.g., \method{SPLADE} and \method{uniCOIL}) and the document expansion models (e.g., \method{TILDE}) all use the language model as backbones, all the components in the learned sparse retrieval systems, including the acquired inverted index, are influenced by the corpus-specific vocabulary (\method{CSV}).}
\label{fig:workflow}
\end{figure}

This section introduces the notion of {\bf Corpus-Specific Vocabularies} (\method{CSV})
and shows how it can be incorporated into different aspects of the overall training procedures of sparse retrieval models: vocabulary selection,
pre-training, document expansion, and model training (see Figure~\ref{fig:workflow} for an overview).
We find that \method{CSV} provides greater coverage of query terms, can be easily incorporated into training procedure of different models, and better correspond to the actual usage of the vocabulary entries in the downstream ranking task inside the inverted index. 

\subsection{Vocabulary Selection}

Before the advent of learned sparse models based on language models, it was common practice for an inverted index to contain lists for all unique tokens in the target corpus.
Standard text collections such as \govtwo{} contain 25 million unique tokens~\cite{ov14-sigir} comprising parsing errors, named entities, numbers, etc. Indexing these unique tokens has the benefit of being able to precisely (and efficiently) retrieve documents containing rare tokens, a key benefit of sparse retrieval models over alternative dense retrieval systems. 

On the contrary, due to computational restrictions (parameter memory usage, softmax inefficiencies among others) associated with Transformers, it is common practice to limit the number of unique tokens fed into such models to only tens of thousands (e.g., 30,000 in the case of standard BERT~\cite{Devlin-etal-2019-bert}). Algorithms such as byte-pair encoding (BPE)~\cite{bpe} or WordPiece~\cite{wu2016google} have been developed to tokenize text into sub-word units, to minimize the occurrence of out-of-vocabulary tokens during text processing with such limited vocabulary sizes. These vocabulary size restrictions are generally non-problematic, as these sub-word tokens are represented in the context of word sequences during standard NLP tasks such as machine translation.

However, in the context of sparse retrieval models such as \uniCOIL, this
contextualization of sub-word tokens only takes place at training time. At retrieval time when using a standard inverted index, each token is processed in isolation. 


We propose to adjust the vocabulary used in sparse models such as \uniCOIL~(and the underlying language model) to better account for this mismatch. For simplicity,
we train WordPiece tokenizers on our target corpus with varying, larger vocabulary sizes.
While vocabulary selection could be enhanced by incorporating other signals such as query logs and term frequency counts into the learning process, we seek to isolate the effect of vocabulary size in this work and leave these extensions to future work. We refer to this process as leveraging Corpus-Specific Vocabularies (\method{CSV}).
In this work, we specifically experiment with vocabulary sizes of $30,000$, $100,000$, and $300,000$. As we will show in detail in Section~\ref{sec:experiments}, increasing
vocabulary size has positive effects on both retrieval quality and runtime latency.



\subsection{Pre-training Objectives}

Since our model employs a different vocabulary from BERT, we cannot use pre-trained BERT checkpoints. BERT is pre-trained with two objectives in mind: Masked Language Modeling (MLM) and Next Sentence Prediction (NSP)~\cite{Devlin-etal-2019-bert}. Pre-training is usually performed on the BooksCorpus (800M words) and English Wikipedia (2,500M words) datasets. Models such as coCondenser~\cite{cocondenser} and \SPLADE~\cite{Formal-2021-spladev2,Formal-etal-2021-splade,lassance2022efficiency} begin with a pre-trained BERT checkpoint and undergo further pre-training on the retrieval corpus, which is sometimes referred to as ``middle-training''~\cite{lassance2022efficiency}. 

In this study, we \emph{bypass} the pre-training step on large, out-of-domain text corpora (e.g. BooksCorpus and Wikipedia) and \emph{only} pre-train on the target retrieval corpus. We make this choice due to (1)~the aim of mitigating the cost and environmental impact associated with pre-training multiple LMs using multiple different vocabularies on large corpora, and (2)~empirical evidence suggesting that LMs pretrained from scratch on the retrieval corpus exhibit improved effectiveness in retrieval tasks~\cite{ir-pretrain}.

Note that pre-training with large vocabulary sizes can be computationally expensive but methods such as hierarchical or sampled softmax are standard, drop-in replacements for softmax cross entropy which improves scalability with regard to the number of classes (vocabulary entries). 

\subsection{Sparse Retrieval Models}

We purpose to use \method{CSV} to enhance the efficiency and effectiveness of sparse retrieval models that rely on the underlying language model vocabulary as their index vocabulary. To demonstrate this, we leverage \uniCOIL~\cite{coil,arxiv21lm} and \SPLADE~\cite{Formal-2021-spladev2,Formal-etal-2021-splade} as examples. \uniCOIL~assigns an impact score to each query and passage token, and discards tokens with a non-positive impact. It relies on an additional model for document expansion. 

\SPLADE, on the other hand, projects queries and passages into $|V|$-dimensional embeddings, where $|V|$ is the vocabulary size of the underlying LM, and calculates the matching score based on the dot product of these embeddings. To ``sparsify'' these dense representations for efficiency, \SPLADE{} employs FLOPS regularizers~\cite{paria2020minimizing} to restrict the number of tokens with non-zero weights. Unlike \uniCOIL{}, \SPLADE{} performs query and document expansion automatically, without the need for prior corpus expansion. Because query expansion significantly influences retrieval efficiency, which is not the focus of this study, we limit our experiments to \SPLADE-\method{DOC} proposed in \method{SPLADE-v2}~\cite{Formal-2021-spladev2}. \SPLADE-\method{DOC} only performs token weighting and expansion on the passage side and assigns uniform weight to query tokens without expansion. Note that changing vocabulary size affects how \SPLADE~ should be regularized, as the FLOPS loss is calculated by \emph{summing} the square of a token's average absolute weight in a mini-batch \emph{across the vocabulary}.

\subsection{Document Expansion}

We use \TILDE~\cite{tildev1} for document expansion with \uniCOIL{}, as it requires fewer resources for both training and inference and provides comparable performance to \DocTQuery~\cite{doct5query}. 
\TILDE{} is initially trained using labeled relevant query-passage pairs. However, the shallow labeling of \msmarcovone{}~\cite{sparse-labels} and the presence of false negatives~\cite{false-negs} make this approach restricted. To enhance \TILDE{} for document expansion, we propose to aggregate the rankings of 12 dense rankers~\cite{hard-negatives} using the Borda Count~\cite{aslam2001models} into a consolidated ranking, and then use the top 10 passages from this ranking as training signals to train \TILDE{}. We choose the Borda Count for ranking aggregation due to its simplicity. Note that \TILDE{} can also leverage CSV as it predicts additional document tokens over the underlying LM vocabulary space which we adjust and fine-tune to the target corpus. We refer to this approach as {\bf Corpus-Specific Document Expansion} leveraging augmentation (\TILDEACSV) as both the vocabulary used during expansion and the underlying expansion model fit directly to the target corpus. The positive benefits of these enhancements will be explored in detail in Section~\ref{sec:experiments}.

\subsection{Distillation-based Training}

Training a student model using the outputs of a trained teacher ranking model as training signals can considerably enhance learning outcomes~\cite{colbertv2,Formal-2021-spladev2,lassance2022efficiency}. We use KL Divergence~\cite{kullback1951information,hinton2015distilling} as the training loss and a standard cross-encoder~\cite{nogueira2019passage} as the teacher to train the \uniCOIL{} and \SPLADE\method{-DOC} models as suggested by~\citet{lassance2022efficiency}.

\vspace{0.25cm}
Overall we seek to apply \method{CSV} to a variety of state-of-the-art techniques, showing they are broadly applicable and generalize to different approaches, which we will show in Section~\ref{sec:experiments}.
\section{Experiments}
\label{sec:experiments}

\subsection{Setup}

\myparagraph{Datasets} We use the \method{MS MARCO v1} (referred to as {\msm} in tables; $8.8M$ passages) and \method{MS MARCO v2} ($138M$ passages) collections. For evaluation, we mainly use the 6,980 queries from the \method{MS MARCO v1 Dev} set. We also use the test queries of TREC 2019~\cite{trec2019} and 2020~\cite{trec2020} from the TREC Deep Learning track.

\myparagraph{Latency experiments} We use the PISA engine~\cite{pisa19-osirrc} which substantially outperforms Lucene in terms of space usage and runtime efficiency for retrieval over learned sparse indexes {\cite{mtl21-arxiv}}. 
We use the state-of-the-art \method{BlockMaxWand}~\cite{ds11-sigir} dynamic-pruning based query algorithm. All our indexes leverage recursive-graph-bisection~\cite{dk+16-kdd,fast-rgb,mm+19-ecir} to optimize efficiency.
Our code and experimental setup is available at \url{https://github.com/PxYu/CSV-for-LSR-ECIR24}.
We report latency as mean retrieval time (MRT) in ms averaged over 5 runs.

\myparagraph{Hardwares} All models are trained on $8\times A100$ GPUs, whereas our latency experiments are performed on an Intel Xeon 8375C CPU in single-threaded execution mode.

\myparagraph{Models and Baselines}

While adjusting the underlying LM and pretraining on the target corpus is general, we focus specifically on the impact on sparse retrieval models. Here we experiment with \SPLADE~ (abbreviated as \SPL) and \uniCOIL~(abbreviated as \uCOIL). For each, we experiment with vocabulary sizes of $30,000$, $100,000$, and $300,000$. We pre-train our \method{CSV} models using the MLM objective (with a 15\% masking probability) on \msmarcovone{} (497M words) (and \method{MS MARCO v2} as indicated) for 10 epochs. \uniCOIL{} models can also leverage \TILDEACSV{} (abbreviated as \TILDEA) document expansion by first training \TILDEACSV{} with query likelihood and document likelihood for 5 epochs and take the top-200 tokens predicted in the document likelihood distribution as expansion, ignoring stop-words, sub-words and tokens in the original passage~\cite{tildev1}. We also train a distillation based version of \uniCOIL~(abbreviated \method{\uCOIL-D}) as discussed in Section~\ref{sec:csv}.

As baselines, we retrain a \uniCOIL~ model with the standard BERT vocabulary (to compare to our $30,000$ vocabulary models, to which it has similar vocabulary size). We additionally report numbers for existing \method{BT-SPLADE-L} model~\cite{lassance2022efficiency} as competitive efficient and effective baselines. We also compare to standard \method{BM25}, \method{DocT5Query} and \method{ColBERTv2}~\cite{colbertv2} baselines.


\subsection{Vocabulary Selection and Index Statistics}

First, we explore the effect of vocabulary selection on different index and query statistics without document expansion.
Table~\ref{tab:qry-vocab-selection} shows query statistics for a \uniCOIL~index (trained only on \msmarcovone{}; no document expansion) with different vocabularies. We observe that the mean number of query tokens decreases as the vocabulary size increases. The number of queries where any sub-word token (compared to only having full word tokens) is present decreases from 48\% for the default BERT vocabulary to 35\% for our custom vocabulary of the same size. Larger vocabularies further decrease the number of queries containing sub-word tokens to 11\% and 2\% respectively. We also observe that passage length, postings per query, and mean retrieval time (MRT) decrease as the vocabulary size increases. Also, note that a custom vocabulary with 30k tokens outperforms the regular \method{BERT-30k} vocabulary on all metrics. Overall, \method{CSV-300k} is 20\% faster compared to a standard \method{BERT-30k}
based \uniCOIL{} model.


\begin{table}[t]
\caption{Mean query length ($|Q|$), percentage of split queries, passage length ($|D|$, in terms of tokens), postings per query, and MRT. 
Metrics are derived from \uniCOIL~models without document expansion. \label{tab:qry-vocab-selection}}
\centering
\begin{tabular}{lccccc}
\toprule
Vocab &  $|Q|$ & \%Split Qrys & $|D|$ & Postings & MRT \\ 
\midrule
\method{BERT-30K} & 7.02 & 48.27\% &  47.6 & \num{6482729} & 22.88 \\
\method{CSV-30K} & 6.68 & 35.50\% & 46.2 & \num{6207331} & 19.70 \\ 
\method{CSV-100K} & 6.29 & 11.36\% &  42.5 & \num{5502811} & 18.66 \\
\method{CSV-300K} & 6.17 & 2.36\% &  41.3 & \num{5118462} & 18.62 \\


\bottomrule
\end{tabular}
\end{table}

\subsection{Retrieval Quality}

\begin{table}[t]
\caption{MRR and MRT for different pre-training, document-expansion and \uniCOIL{} training
objectives.
~\label{tab:quality-speed-1}}
\centering
\begin{tabular}{ccccccc}
\toprule
\# & Vocab & Pretrain & D. Exp. & Model & \multicolumn{1}{c}{MRR} & \multicolumn{1}{c}{MRT}  \\ 
\midrule
1 & \method{\footnotesize BERT} & \method{\footnotesize BERT} & \method{\footnotesize TILDE} & \method{\footnotesize uCOIL}  & 0.354 & 33.88  \\ 
2 &\method{\footnotesize BERT} & \msm{} & \method{\footnotesize TILDE} & \method{\footnotesize uCOIL}  & 0.343 & 29.29 \\
\midrule
3 &\method{\footnotesize CSV-100K} & \msm{} & - & \method{\footnotesize uCOIL} & 0.332 & 18.66 \\ 
4 &\method{\footnotesize CSV-100K} & \msm{} & \method{\footnotesize TILDE} & \method{\footnotesize uCOIL} & 0.353 & 22.65 \\   
5 &\method{\footnotesize CSV-100K} & \msm{}\method{+v2} & \method{\footnotesize TILDE} & \method{\footnotesize uCOIL} & 0.370 &  22.45 \\

6 &\method{\footnotesize CSV-100K} & \msm{}\method{+v2} & \method{\footnotesize TILDE-A} & \method{\footnotesize uCOIL} & 0.376  & 19.88 \\ 
7 &\method{\footnotesize CSV-100K} &  \msm{}\method{+v2} & \method{\footnotesize TILDE-A} & \method{\footnotesize uCOIL-D} & 0.391 & 18.85 \\ 

\bottomrule
\end{tabular}
\end{table}

Table~\ref{tab:quality-speed-1} explores pretraining, document expansion, and model distillation. Note that we omit certain configurations and metrics that do not provide additional insights to simplify presentation. 
Rows \#1 and \#2 represent reproduced standard baselines for reference.

First, comparing \method{CSV-100k} with no document expansion (row \#3) and \method{CSV-100k} with standard \TILDE{} expansion (row \#4), we observe that latency increases but retrieval quality improves.
Both pre-training on \method{MS MARCO v2} (row \#5) and augmented document expansion (\TILDEA) improve retrieval quality while remaining latency neutral or improving latency.
Finally, we replace regular \uniCOIL{} with a version trained with distillation (\method{uCOIL-D}, row \#7).
\method{\uCOIL{}-D} provides the best retrieval quality ($0.391$ MRR). Note that this is competitive to state-of-the-art late interaction models such as \method{ColBERTv2}~\cite{colbertv2} (0.397 MRR on the same task). In subsequent experiments,
we restrict our analysis and presentation to pre-training on \method{MSM+v2}, expansion using \TILDEA{} and \method{\uCOIL{}-D}.

Table~\ref{tab:quality-speed-2} shows the effect of increasing vocabulary sizes on \uniCOIL{} based models. No substantial difference between the \method{CSV-100k} (row \#10) and \method{CSV-300k}  (rows \#11-\#13) can be observed, which indicates that $100k$ is a sufficient vocabulary size for \method{MS MARCO v1}. We also observe that latency {\it increases} for \method{CSV-300k} (row \#13) compared to \method{CSV-30k} (row \#9), which is contrary to the numbers reported in Table~\ref{tab:qry-vocab-selection}. We find \TILDEA{} expansion increases document size substantially with larger vocabulary size ($90.71$ extra tokens on average 
for \method{CSV-300k}, $41.63$ for \method{CSV-30k}). This increase counteracts the decrease in postings list lengths we obtained through increasing vocabulary size. Adjusting the \TILDEA{} hyperparameter to only expanding with the top-40/50 tokens, in rows \#10 - \#12 we see latency in line with \method{CSV-30k} for larger vocabularies while showing a negligible improvement in retrieval performance. In summary, for \method{MS MARCO v1}, leveraging a custom vocabulary (rows \#9 compared to \#8) is more important to improving retrieval quality compared to increasing vocabulary size, which however has a positive impact on latency.

Table~\ref{tab:quality-speed-3} shows the effect of corpus-specific vocabulary in different sizes on \SPLADE. Similarly, as with \uniCOIL{} (row \#8 and \#9), the \method{CSV} model (row \#16) outperforms the model with BERT vocabulary in similar size (row \#17) in terms of MRR and MRT.
Again, retrieval quality does not increase with larger vocabulary sizes, however, the \method{CSV-300k} version (rows \#20 and \#21) is roughly 40\% faster than the comparable \method{\SPL{}-DOC} baseline (row \#15). This effect is related to the FLOPS sparsity regularization leveraged by \SPLADE{} interacting with vocabulary size. Experimenting with different regularization strengths ($\lambda_d$) while trying to keep MRR roughly constant (rows \#17 - \#21), we find larger vocabularies ($300k$) result in improved retrieval speed ($13ms$ vs $25ms$).


\begin{table}[t]
\centering
\caption{MRR and MRT for different custom vocabulary sizes and document expansion limits. \# Kept Tokens refers to the number of expansion tokens provided by \TILDEA~ that are actually used for document expansion. It acts as a hyperparameter to control the balance between effectiveness and efficiency under the same vocabulary. ~\label{tab:quality-speed-2}}
\begin{tabular}{cccccc}
\toprule
\# & Vocab & \# Tilde Tokens & \# Kept Tokens & \multicolumn{1}{c}{MRR} & \multicolumn{1}{c}{MRT} \\ 
\midrule
8 &\method{\footnotesize BERT-30K}  & 37.5 & 37.5 & 0.379 & 20.00 \\ 
9 &\method{\footnotesize CSV-30K}  & 41.6 & 40.0 & 0.389 & 18.69 \\ 
10 &\method{\footnotesize CSV-100K} & 46.6 & 40.0 & 0.389 & 17.24  \\ 
11 & \method{\footnotesize CSV-300K} & 90.7 & 40.0 & 0.388 & {\bf 17.06}  \\ 
12 & \method{\footnotesize CSV-300K} & 90.7 & 50.0 & {0.391} & 18.72  \\ 
13 &\method{\footnotesize CSV-300K} & 90.7 & 90.7 & {\bf 0.392} & 22.08  \\ 

\bottomrule
\end{tabular}

\end{table}


Similarly, Table~\ref{tab:trec-qrys} shows that our improvements  also transfer to the \method{TREC} query sets. While standard \method{BM25} and \method{DocT5Query} are still faster, \method{CSV} reduces mean latency relative to regular \uniCOIL{} by 50\% ($17.63ms$ vs $33.73ms$) and improves over state-of-the-art \method{BT-SPLADE-L} method.
We conduct Bonferroni corrected pairwise t-tests, and report significance with $p < 0.05$.

\begin{table}[t]
\centering
\begin{threeparttable}
\caption{MRR and MRT for several \SPLADE{} based methods. ~\label{tab:quality-speed-3}}
\centering
\begin{tabular}{cccccc}
\toprule
\# & Method & Vocab & $\lambda_d$ & \multicolumn{1}{c}{MRR} & \multicolumn{1}{c}{MRT}  \\ 
\midrule
14 & \multicolumn{3}{c}{\method{\footnotesize BT-SPLADE-L}~\cite{lassance2022efficiency}} & 0.380  & 27.62  \\ 
15\tnote{$\dagger$} &\method{\footnotesize SPL-DOC} & \method{BERT-30K}   &  0.008  & 0.347 & 21.03 \\
\midrule
16 &\method{\footnotesize \SPL{}-DOC} & \method{BERT-30K}  & 0.008 & 0.339 & 27.28 \\
17 &\method{\footnotesize \SPL{}-DOC} & \method{CSV-30K}  & 0.008 & 0.356 & 25.39 \\
18 &\method{\footnotesize \SPL{}-DOC} & \method{CSV-100K} & 0.009 & 0.358 & 21.65  \\
19 &\method{\footnotesize \SPL{}-DOC} & \method{CSV-300K} & 0.006 & {\bf 0.359 } & 18.47  \\
20 &\method{\footnotesize \SPL{}-DOC} & \method{CSV-300K} & 0.007 & 0.357 & {\bf 13.12 } \\
21 &\method{\footnotesize \SPL{}-DOC} & \method{CSV-300K} & 0.008 & 0.354 & 14.21 \\
\bottomrule
\end{tabular}
\begin{tablenotes}
\item[$\dagger$] This is initialized with a DistilBERT model that is further pretrained on MSMARCO using MLM+FLOPS~\cite{lassance2022efficiency}. In comparison, row \#16 is initialized with a BERT model that is only pretrained on MSMARCO v2 using MLM.
\end{tablenotes}
\end{threeparttable}
\end{table}



\begin{table}[t]
\caption{nDCG@10 and MRT, for \method{TREC 19\&20} queries. The symbol $\triangledown$ denotes a sig. difference viz. \uCOIL\method{-D-CSV-300K (\#13)}. \label{tab:trec-qrys}}
\centering
\begin{tabular}{lcccc}
\toprule
Strategy & \multicolumn{2}{c}{TREC 2019} & \multicolumn{2}{c}{TREC 2020}  \\ 
\cmidrule(lr){2-3} \cmidrule(lr){4-5} 
&\multicolumn{1}{c}{nDCG} & \multicolumn{1}{c}{MRT} & \multicolumn{1}{c}{nDCG} & \multicolumn{1}{c}{MRT}    \\
\midrule
\method{BM25} & 0.501$^\triangledown$& 4.93 &  0.487$^\triangledown$ & 7.94\\
\method{DocT5Query} & 0.643$^\triangledown$ & 4.87 &  0.607$^\triangledown$ & 7.80 \\
\method{UniCOIL-TILDE} & 0.660$^\triangledown$& 31.59 & 0.647$^\triangledown$ & 33.73 \\ 
\method{BT-SPLADE-L}   & 0.703  & 26.91 & 0.698 & 27.60 \\
\midrule
\method{\footnotesize \uCOIL-D-CSV-100K (\#10)}  & 0.718  & 15.00 & 0.706 & 18.01\\
\method{\footnotesize \uCOIL-D-CSV-300K (\#11)} & 0.722 & {\bf 14.46} & 0.708 & {\bf 17.63} \\
\method{\footnotesize \uCOIL-D-CSV-300K (\#13)} & {\bf 0.729}  & 17.55 & {\bf 0.728} & 22.13 \\
\bottomrule
\end{tabular}
\end{table}


\subsection{Query Latency}

\label{sec:analysis}


\begin{figure}[t]
\begin{center}
\includegraphics[width=0.90\columnwidth]{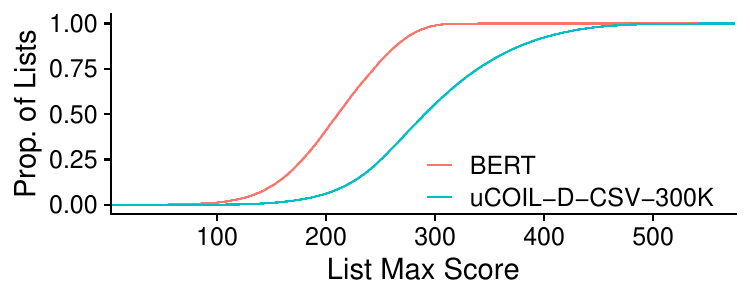}
\end{center}
\caption{Cumulative distribution of lists that have list max scores higher than a given value. \method{BERT} displaying
less skew in list max scores which negatively affects performance. \label{fig:impactdist}}
\end{figure}

Previous experiments show that \method{CSV} with both 100k and 300k tokens substantially reduces the latency of existing approaches. For example, standard \uniCOIL~with BERT vocabulary (row 1; Table~\ref{tab:quality-speed-1}) exhibit a mean response time of $33.88$ms, whereas our fastest method \uniCOIL~based method reduces mean response time to $17.06$ms (row 11; Table~\ref{tab:quality-speed-2}), a 50\% reduction. Similarly, \SPLADE{} enhanced by \method{CSV} exhibits similar latency improvements (see Table~\ref{tab:quality-speed-3}). For comparison, \method{ColBERTv2}~\cite{colbertv2} accelerated by \method{PLAID}~\cite{plaid} provides similar effectiveness but is substantially slower ($185$ms single CPU; not run by us).

We observe that \method{CSV-300K} results in more lists (due to having a larger vocabulary) with larger list max scores referring to maximum score a term is assigned in any document in the collection as shown in Figure~\ref{fig:impactdist}. This also creates more skewed list max score distribution (the score ``band'' in Figure~\ref{fig:impactdist} is more narrow for the standard BERT vocabulary) which is essential as pruning algorithms use list max scores to skip over low-scoring documents~\cite{ds11-sigir}. 

This has a direct effect on runtime performance which can be observed in the run-time statistics of the \method{MaxScore} algorithm shown in Table~\ref{tab:maxscorestats}. Note that methods \method{uniCOIL} and \method{BT-SPLADE-L} which leverage smaller vocabularies score substantially more documents. This would be especially impactful in the case where a non trivial scoring function (e.g. scoring discovered documents with a more expensive secondary model as in proposed by~\citet{mmst2022gt}) is used to score documents.

Interestingly, operation \method{Insert} which counts the number of insertions into the final top-$k$ result heap during processing are similar. While more documents are scored, the amount of documents inserted into the resulting heap stays similar. This is an artifact of list max scores (plotted in Figure~\ref{fig:impactdist}) being used to determine if a document should be scored and larger vocabularies provide more fine-grained ``decision boundaries'' as fewer high and low-scoring terms are conflated into a single vocabulary entry.  

\begin{table}[t]
\caption{Query processing statistics (avg per query) for \method{MaxScore} and three index varieties. \label{tab:maxscorestats}}
\centering
\begin{tabular}{lcccc}
\toprule
Strategy & \multicolumn{1}{c}{SCORE} & \multicolumn{1}{c}{INSERT} & \multicolumn{1}{c}{NEXT} & \multicolumn{1}{c}{NEXT-GEQ}  \\ 
\midrule
\uniCOIL & \num{7004022} & \num{301650} & \num{6575301} & \num{2048790} \\ 
\method{uCOIL-D-CSV-300K} & \num{4836943} & \num{450567} & \num{4554723} & \num{1336857} \\
\method{BT-SPLADE-L} & \num{6400831} & \num{556829} & \num{6110824} & \num{1175790}  \\
\bottomrule
\end{tabular}
\end{table}

\subsection{Pre-training Cost and Model Size} 
Not using existing LM checkpoints requires more time and resources for pre-training. 
We pretrain each LM on \msmarcovone{} for 10 epochs with MLM. For experiments with \uniCOIL{}, we train \TILDE{} for 5 epochs for document expansion, and train \uniCOIL{}~for 5 epochs on the expanded corpus for retrieval. \SPL{}\method{-DOC} is trained for 50k iterations using our pre-trained LM. We spend 4-13 hours pretraining LMs of different sizes due to the computational overhead of larger vocabulary sizes. Our pretraining does not currently leverage standard sampling/hierarchical softmax strategies used to deal with a large number of categories, which increases the cost. Increasing vocabulary size also increases model parameter size from 109M to 316M, similar to \method{BERT-large} with a 30k vocabulary. We experimented with pre-training \method{BERT-large} on our corpus to obtain a baseline with similar parameter size, but found that the \method{MS MARCO} corpora were too small to pre-train a model of this size. 
Note that search-specific pre-training tasks such as coCondenser~\cite{cocondenser} provide orthogonal benefits to vocabulary changes. We leave exploring potential interactions of these techniques to future work.

\section{Conclusion and Future Work}

We demonstrate that corpus-specific vocabularies are effective at improving both retrieval quality and query latency of learned sparse retrieval systems. They are simple yet effective and can be applied to a variety of different modeling types. 

We believe there is a large body of future work exploring the effect of the vocabulary on sparse retrieval models. Promising directions are developing more sophisticated vocabulary selection strategies and training and document expansion strategies that take underlying inverted index-based retrieval into account when assigning term weights.

%
%

\bibliographystyle{splncs04nat}
\bibliography{strings-long,main}
\end{document}